\documentstyle[12pt]{article}
\newcommand{\bra}[1]{\langle #1 |}
\newcommand{\ket}[1]{| #1 \rangle}
\def\<{{\langle}}
\def\>{{\rangle}}

\def\vx{{\vec x}}

\def\vp{{\vec p}}
\def\vP{{\vec P}}
\def\sp{{\sf p}}
\def\sP{{\sf P}}
\def\sJ{{\sf J}}

\def\sx{{\sf x}}
\def\sy{{\sf y}}
\def\sh{{\sf h}}
\def\sX{{\sf X}}

\def\sE{{\sf E}}

\def\S{{\cal S}}
\def\H{{\cal H}}

\def\sV{{\sf V}}
\def\U{{\cal U}}

\def\be{\begin{equation}}
\def\ee{\end{equation}}
\def\bea{\begin{eqnarray}}
\def\eea{\end{eqnarray}}
\def\half{{\textstyle{1\over 2}}}
\def\fourth{{\textstyle{1\over 4}}}
\def\thalf{{\textstyle{3\over 2}}}
\date{ }
\begin{document}

\title{Relativistic Quantum Dynamics of Many-Body Systems}

\author{F. Coester\\
\small
Physics Division,Argonne National Laboratory, Argonne, IL
60439,  USA\\ \small E-mail:
coester@anl.gov\\
W. N. Polyzou\\
\small
Department of Physics and Astronomy, 
University of Iowa, Iowa City IA 52242,USA\\
\small E-mail: polyzou@uiowa.edu}

\maketitle

\abstract{
Relativistic quantum dynamics requires a unitary representation
of the Poincar\'e group on the Hilbert space of states.
The dynamics of many-body systems must satisfy cluster separability
requirements. In this paper we formulate an abstract framework
of four-dimensional Euclidean Green functions
that can be used to construct relativistic quantum dynamics of
$N$-particle systems  consistent with these requirements. This approach
should be useful in bridging the gap between few-body dynamics based on
phenomenological mass operators and on quantum field theory. }

\section{Introduction}

The superposition principle and  the space-time symmetry
are realized in relativistic quantum mechanics by a Hilbert
space of states with a unitary representation of the inhomogeneous 
Lorentz group (Poincar\'e group).\cite{wigner} Various representations of single
particle states are well known.\cite{haag_weinberg} Since the components of the
four-momentum $p$ are constrained by the mass, $p^2=-m^2$, there is a choice
of convenient independent   momentum variables in the wave function:
For instance the components $\vp$ orthogonal to some fixed time-like vector,
or the components $p^+,p_\perp$ orthogonal to some null-vector.
For particles with spin the functions representing state vectors are 
functions of spin variables undergoing Wigner rotations.
All these representations are equivalent. States of noninteracting particles are
represented by tensor products of single particle states.

In quantum mechanics the Hilbert space of state vectors is the same for 
free and interacting particles. The interactions are implemented by modifications
of the Poincar\'e generators. Following Bakamjian and Thomas\cite{bak}
this has been done modifying the mass operator, leaving the spin operator
independent  of interactions. The Poincar\'e generators obtain as functions of
 kinematic components  of the four
momentum operator and canonically conjugate positions, 
the mass operator, the spin operator. The choice of these kinematic components
determines the ``form of dynamics''.\cite{dirac} The principal difficulty in this approach is
the realization of cluster separability.\cite{cluster1} The properties
of any isolated cluster of particles should not depend on the
presence or absence of other clusters. The solution  involves 
the recursive construction of appropriate many-body interactions
in the many-body mass operators.\cite{cluster2} There are no theorems  defining
minimal many-body interactions.

Alternatively, single-particle states  can be represented by equivalence classes of
covariant functions  of the four-momentum  with the positive semi-definite 
inner product measure $d\mu(p):=d^4p\delta(p^2+m^2)\theta(p^0)$.
For particles with spin the functions representing state vectors are Lorentz covariant
functions of spinor indices with a semi-definite inner product measure.
Starting from kinematically covariant functions representing multi-particle states, interactions 
can be introduced modifying  the semi-definite inner product measure.\cite{constraint-wnp}

For  free fields  (free particles) Poincar\'e generators obtain by integration
of the energy-momentum tensor over a three-dimensional manifold in the 
Minkowski space.\cite{schwinger} Interactions in the energy-momentum
tensor require local commutation relations and infinitely many
degrees of freedom. The action of these Poincar\'e generators on Fock-space vectors
produces linear functionals over the Fock space, not vectors in Fock-space.
The Hilbert spaces of free and interacting fields are necessarily inequivalent.
 
Minkowski-Green functions are defined by vacuum expectation values of time-ordered
products  of local renormalized Heisenberg fields. Using assumed spectral
properties of the intermediate states and the asymptotic properties of 
the field operators there are simple relations to observable bound-state masses,
scattering  amplitudes and form factors. The principal problem is to establish
a quantitative relation of ``approximations'' to  a theory local operators.

A central feature in the formulation of relativistic quantum theory
is the absence of finite-dimensional unitary representations  of the Lorentz group, 
$O(1,3)$. This is the reason the representation of states by Lorentz covariant
functions  requires a semidefinite inner product measure.
However, the Lorentz group is related, by complexification, to the 
orthogonal group in four dimensions, $O(4)$ which does have finite dimensional unitary
representations. The  unitary representations of the real Euclidean group $E(4)$
together with invariant Green operators are useful in the formulation
of Poincar\'e invariant dynamics. This connection has been
exploited extensively in quantum field theory. The equivalence of the Wightman axioms
 with the Osterwalder-Schrader axioms\cite{osterwalder}
establishes that Euclidean Green functions (Schwinger functions) satisfying the 
Osterwalder-Schrader axioms imply the  existence of field operators.
In the context of Lagrangean field theory the Schwinger functions obtain as
moments of the functional measure defined by the Lagrangean. In the
Euclidean formulation the locality axiom is independent of the axioms  which
establish  unitary Poincar\'e representations with the appropriate spectral 
properties. This feature is essential for the formulation of 
Poincar\'e covariant dynamics finite many-body systems.

The purpose of this paper is to explore the formulation of 
relativistic quantum dynamics based on Euclidean invariant Green functions.
Ordinary quantum mechanics provides some heuristic indications. The Hilbert space, $\H$
of states of
$N$ particles is independent of the dynamics. It is the same for free and interacting
particles. The dynamics is specified by the invariant Casimir  Hamiltonian,
$
\sh:= H-\vP\,^2/2M
$
or the resolvent  operator $G(\sE):=1/(\sh-i\sE)$. Interactions are introduced by
invariant  additions to the Casimir Hamiltonian
$
\sh =\sh_0+V,$ or $ G(\sE)^{-1}= G_0(\sE)^{-1}+ V.
$
While the approach explored here should be applicable to
many qualitatively different physical systems and illuminate relations to
quantum field theory the focus of
 this exploratory study is limited 
systems with a fixed number of particles, for instance nucleons. 
The abstract framework of the dynamics to be explored is formulated in Section 2. 
Section 3 provides a realization for single particles. Two-body interactions are
formulated in Section 4.
The many-body dynamics
with the realization of cluster separability is discussed in Sec. 5.

\section{The Abstract Framework}
\subsection{The Auxiliary Hilbert Space.}
Since there are no finite dimensional unitary representations of
the Lorentz group $SO(1,3)$ we assume the representatives of 
physical states to be a subset of a larger space, subject to 
the following assumptions. 

\begin{description}

\item{H1.} Physical states are a linear subset of vectors $\Psi$ in  an
auxiliary Hilbert space
${\cal H}_a$ with the norm
$\Vert \Psi\Vert_a^2 =\langle\Psi,\Psi\rangle$. 

\item{H2.} There is 
a unitary representation $U({\cal R},a)$ of the Euclidean group 
$E(4)$ with ${\cal R}\in O(4)$ on the  Hilbert space $\H_a$.
 The self-adjoint generators of $E(4)$ are
denoted by ${\sf P}^\mu$ and  ${\sf J}^{\mu\nu}=-{\sf J}^{\nu\mu}$.
\item{H3.}  There is a self-adjoint, unitary operator $\Theta$, on $\H_a$,
which is invariant under the 3-dimensional Euclidean subgroup
\be
[\Theta,\sP^k]=0\; ,\qquad\qquad [\Theta,\sJ^{i k}]=0\; ,
\ee
and satisfies
\be
\Theta\sP^0=- \sP^0\Theta\; ,\qquad
\Theta \sJ^{0k}=- \sJ^{0k}\Theta\; .
\ee
 \end{description}
The two Casimir operators $\vec j_\pm^2$ of $O(4)$ are functions of the generators
$\sJ^{\mu\nu}$,\cite{schweber}
\be
 j^k_\pm:={1\over \sqrt{2}}\left( \sum_{\mu<\nu}\epsilon^{0k\mu\nu}\sJ^{\mu\nu}\pm 
\sJ^{0k}\right)\; .
\ee
The spectra of the Casimir operators are $\sigma(\vec j_\pm^2)= s_\pm(s_\pm+1)$
with non-negative integer or half-odd integer values of $s_\pm$.

The operators
\be
P^0:=i\sP^0\;,\qquad P^k:=\sP^k\;,\qquad
J^{0k}:=i\sJ^{0k}\;,\qquad  J^{ik}:=\sJ^{ik}
\label{HPG}
\ee
satisfy the Poincar\'e Lie algebra, and the spectrum of $P^2$ is the real line,
\newline
$-\infty<\sigma(P^2)<\infty$. The operators
\be
\exp(iP^0t)\equiv\exp(-\sP^0t)\qquad \mbox{and}\qquad \exp(iJ^{0k}\chi)\equiv\exp(-\sJ^{0k}\chi)
\ee
are self-adjoint. 
Together with the unitary representations of the 3-dimensional 
Euclidean group they define a non-unitary representation of the Poincar\'e group.
The inner product 
$(\Psi,\Psi):=\<\Theta\Psi,\Psi\>\equiv \<\Psi,\Theta\Psi\>$  
defines a pseudo-Hilbert space\cite{bogol} of the vectors in $\H_a$. 
The inner product $(\Psi,\Psi)$ of the pseudo-Hilbert space is Poincar\'e invariant,
$
(U(\Lambda, a)\Psi,U(\Lambda, a)\Psi)=(\Psi,\Psi).
$

\subsection{Green Operators and the Physical Hilbert Space.}
The representation of physical states and the dynamical properties of the system are
specified by a Green operator,
$G$, with the  following properties.
\begin{description}
\item{G1.}  The Green operator $G$ is a 
 bounded normal operator on $\H_a$ with an inverse defined on a dense set.

\item{G2.} The Green operator $G$  commutes with 
$\sP^\mu$ and
$\sJ^{\mu\nu}$ and hence with $P^\mu,\,J^{\mu\nu}$.

\item{G3.} The operator $\Theta G$ is Hermitean,
$
\Theta G = G^\dagger \Theta\,.
$
\item{G4.}  There is a Poincar\'e invariant linear manifold $\S$ of  vectors
$\Psi\in
\H_a$  that all $\Psi\in \S$ satisfy the inequalities
\be
0\leq (e^{-i\sP^0\tau}\Psi,Ge^{-i\sP^0\tau}
\Psi)\equiv(\Psi,Ge^{-2P^0\tau}\Psi)\leq (\Psi,G \Psi)\; ,\quad \forall\;
\tau \geq 0\; .
\label{SPP}
\ee
\end{description}
By assumption G4 the inner product $(\Psi,G\Psi)$ of vectors in this manifold is
positive semi-definite. 
Physical states are represented by equivalence classes of vectors.
Two vectors $\Psi_a$ and $\Psi_b$ are 
equivalent,
$\Psi_a\sim\Psi_b$ iff 
\be
\Vert \Psi_a-\Psi_b\Vert^2:=
 \left( [\Psi_a-\Psi_b],G [\Psi_a-\Psi_b)]\right)=0\; .
\ee
The physical Hilbert space $\H$ is equipped with
a unitary representation of the Poincar\'e group.
Single-particle states $\Psi_M$ of mass $M$, elementary or composite,
satisfy  $(P^2+M^2)\Psi_M\sim 0$.

\subsection{ Perturbations of Green Operators}
A perturbation $G_0\to G$ may be defined
by
\be
G^{-1}:= G_0^{-1} +\U\; ,
\label{PGU}
\ee
where $\U$ is an  $E(4)$ invariant, pseudo-Hermitean operator 
with domain\newline $D(\U)\supset D(G_0^{-1})$ 
By assumption $\U$ is 
bounded relative to $G_0^{-1}$,
\be
\Vert \U\Psi\Vert_a \leq a \Vert G_0^{-1}\Psi\Vert_a + b\Vert \Psi\Vert_a\; ,
\ee
with $0\leq a<1$ and $0\leq b$.
The
operators $\U G_0$, $G_0\U$ and
 are bounded with a bound less than 1.
It follows that $G^{-1}G_0$, and $G_0G^{-1}$ 
are bounded operators with bounded inverses.

\section{Realization for Single Particles}

The auxiliary Hilbert space $\H_a$ of a spin-zero single particle is realized by square
integrable functions
$\Psi(\sx)$ with $\sx:=\{\sx^0,\sx^1,\sx^2,\sx^3\}$ with the inner product 
\be
\<\Psi,\Psi\>= \int d^4\sx |\Psi(\sx)|^2\; .
\ee
Schwartz functions  $f(\sx)$ are dense in this Hilbert space. 
The involution operator $\Theta$ is defined by
\be
\Theta \Psi(\sx):=\Psi(-\sx^0,\vec\sx)\; .
\ee
The self-adjoint
generators of the real Euclidean group $E(4)$ are
\be
\sP^\mu:={1\over i}{\partial \over \partial \sx^\mu}\; , \qquad\qquad
\sJ^{\mu\nu}:={1\over i}\left(\sx^\mu{\partial \over \partial \sx^\nu}
-\sx^\nu{\partial \over \partial \sx^\mu}\right)\; .
\ee
The associated Poincar\'e generators are then defined by eq. (\ref{HPG}).
The   Green operator is represented by the Green function
$G(\sx-\sx')$ 
\bea
G(\sx-\sx')&:=& \left({1\over 2\pi}\right)^4 \int d^4\sp\, 
{\exp[i\sp(\sx-\sx')]\over
\sp^2+m^2}\cr\cr
&=&\left({1\over 2\pi}\right)^3\int d^3\vp\,{\exp\left(i\vp(\vx-\vx')
-\omega(p)|{\sx'}^0-\sx^0|\right)     \over 2\omega (p)}\; , 
\eea
where $\omega(p):=\sqrt{m^2+\vp\,^2}$.
Schwartz functions  $f(\sx)$ with support restricted to positive
values of $\sx^0$ represent  a linear manifold of vectors in $\H_a$
which satisfies the the requirement G4. 
With the support restriction it follows that
\be
(f,G f)= \<\tilde f,\tilde f\>:=\int {d^3 p\over 2\omega(p)} |\tilde
f(\vp)|^2 = \int d^4\sx\, |\tilde f(\sx)|^2\; ,
\ee
where
\bea
\tilde f(\vp)&:=&
 (2\pi)^{-\thalf}\int d^4\sx e^{-i\vp\cdot \vx}e^{-\omega(p)\sx^0}f(\sx)\; ,\cr\cr
\tilde f(\sx)&:=&(2\pi)^{-\thalf} \int d^3p\, 
e^{i\vp\cdot\vx -\omega(p)\sx^0}\theta(\sx^0)\tilde f(\vp)\; .
\eea

The Hilbert space $\H $ is constructed by the usual procedure of
moding out zero-norm vectors and adding Cauchy sequences.  The equivalence
classes of Schwartz functions
 are dense in  $\H$. Two functions $f_1$
 and $f_2$ are equivalent, $f_1\sim f_2$ iff
\be
\Vert f_1-f_2\Vert =0\; .
\ee
It follows that two functions $f_1$ and $f_2$ are in the same equivalence class,
 $f_1\sim f_2$,\newline iff $\tilde f_1=\tilde f_2$. 

Since
\be
\theta({\sx'}^0)\Theta G(\sx'-\sx)\theta(\sx^0)=
\theta({\sx'}^0)\theta(\sx^0)\left({1\over
2\pi}\right)^3
\int d^3p\, {e^{-\omega(p)
({\sx'}^0+\sx^0)}e^{i\vp\cdot(\vx'-\vx)}\over2\omega(p)}\; .
\ee
it follows that the inner product $(f_a,G\,f_b)$ of functions satisfying the support
condition is manifestly Lorentz invariant,
\be
(f_b,G\,f_a)= \int {d^3p\over \omega(p)}\, \tilde f_b(\vp)^*\tilde f_a(\vp)
=\int d^4p \,\delta(p^2+m^2)\,\theta(p^0)\, f_b(p)^*f_a(p)
\ee
with
\be
f(p):= \int d^4 \sx \exp\left(-i\vp\cdot\vec \sx - p^0 \sx^0\right)\,f(\sx)\; .
\ee
The time evolution evolution $f_a(t):=e^{-iP^0t}f_a$ is given explicitly
by
\be
(f_b,G\,e^{-iP^0t}f_a)= \<\tilde f_b,\tilde f_a(t)\>\; ,
\ee
with $\tilde f_a(t,p):=e^{-i\omega(p) t}\tilde f_a(\vp)$.

For a single spin $1/2$ particle the  Green function is 
\be
G(\sx-\sy) := {1 \over (2\pi)^4} \int d^4 \sp e^{i \sp \cdot (\sx-\sy)} 
\left ( {\sp \cdot \gamma_e+m  \over
\sp^2 + m^2} \right ) \; ,
\label{eq:gspin}
\ee
where the spinor  matrices $\gamma_e := i\beta,\beta  \vec\alpha$
 with $\vec \alpha:=\gamma_5 \vec \sigma $ satisfy
\be
\half\{\gamma_{e\mu},\gamma_{e\nu}\}=-\delta_{\mu\nu}\; .
\ee
The involution operator $\Theta$ must also act on the spinor indices,
\be
(\Theta f) (\sx) := \beta f(-\sx^0, \vec{\sx}).
\ee
As in the case of spin $0$ it is easy to verify positivity of the inner product $(f,G
f)$ for Schwartz functions with support restricted  to positive values of $\sx^0$.
\bea
(f,Gf) &:=& \int d^4\sx d^4\sy f^\dagger (\sx) \Theta G (\sx-\sy) f(\sy)
=\int d^4\sx d^4\sy f^\dagger (\sx)
\beta G (-\sx^0-\sy^0, \vec \sx-\vec \sy) f(\sy) \cr\cr
&=&\int d^3 p  
\tilde f^\dagger (p) 
{\omega (p)  +  \vec \alpha \cdot \vec p + \beta m
\over 2 \omega (p) } \tilde {f} (p) =\int d^4\sx \tilde f^\dagger(\sx)\tilde f(\sx)
\geq 0\;,
\eea
where
\be
\tilde f(\sx):=(2\pi)^{-\thalf} \int d^3p\; e^{i\vp\cdot\vx -\omega(p)\sx^0}\theta(\sx^0)\;
{\omega (p)  +  \vec \alpha \cdot \vec p + \beta m\over 2 \omega (p) }\,\tilde f(\vp)\;.
\ee

\section{Two-Body Dynamics}
The auxiliary Hilbert space $\H_a$ is the tensor product of
the single-particle auxiliary Hilbert spaces. The involution operator $\Theta$
is the tensor product of single particle involution operators.
Schwartz functions
$f(\sx_1,\sx_2)\equiv f(\sX,\sx)$ with $\sX:=\half(\sx_1+\sx_2)$
and $\sx:=\sx_1-\sx_2$ are dense in this Hilbert space. 
For spin $1/2$ particles these functions depend on spinor variables as well.
The $E(4)$ generators are
additive,
\be
\sP^\mu={1\over i}\sum_{n=1}^2 {\partial\over \partial \sx^\mu_n},\; \qquad
\sJ^{\mu\nu}={1\over i}\sum_{n=1}^2\left(\sx_n^\mu{\partial \over \partial \sx_n^\nu}
-\sx_n^\nu{\partial \over \partial
\sx_n^\mu}+\fourth[\gamma_{ne}^\mu,\gamma_{ne}^\nu]\right)\;.
\ee

In general Green operators are  realized  by $E(4)$ invariant  tempered distributions
$G(\sx_1, \sx_2;\sx_2',\sx_1')$.
For free particles the Green function is the product of single particle 
Green functions,
Interactions are added according
to eq. (\ref{PGU}).
A simple example of a nucleon-nucleon interaction is of the form
\bea
\U\Psi(\sX,\sx)&=&
\Bigl\{\sV_S(\sx)  +\gamma_5^{(1)}\gamma_5^{(2)}\sV_P(\sx)
+\sum_{\mu,\nu}
[\gamma_{e\mu}^{(1)},\gamma_{e\nu}^{(1)}][\gamma_{e\mu}^{(2)},\gamma_{e\nu}^{(2)}]
\sV_T(\sx)
\cr\cr
&+&(\gamma_e^{(1)}\!\cdot\!\gamma_e^{(2)})\sV_V(\sx)+
(\gamma_5^{(1)}\gamma_e^{(1)}\!\cdot\!\gamma_e^{(2)}\gamma_5^{(2)})\sV_A(\sx)\Bigr\}
\Psi(\sX,\sx)\; .
\label{BSU}
\eea
Scattering wave functions obtain by the weak time limits
\be
(\Psi_b,G\Omega_{a\pm})= \lim_{t= \pm \infty} (\Psi_b,Ge^{iP^0t}\tilde \Psi_a(t) )
\ee
where $\tilde \Psi_a(t) := \tilde \Psi^{(1)}_a(t)\times \tilde \Psi^{(2)}_a(t)$
is the tensor product of single particle states. \cite{brenig}
The $S$-matrix obtains in the limit
\be
S_{ba}=\lim_{t'= +\infty}\lim_{t= -\infty}
\left(\tilde \Psi_b(t'),G e^{iP^0(t-t')}\tilde \Psi_a(t) \right)\; .
\ee

\section{Many-Body Dynamics}

The auxiliary Hilbert space $\H_a$ is the N-fold  tensor product of
the single-particle auxiliary Hilbert spaces. The involution operator $\Theta$
is the outer product of single particle involution operators.
Schwartz functions,
$f(\sx_1,\dots,\sx_N)$, of $N $points and N spinor indices  are dense in this Hilbert
space. The
$E(4)$ generators are additive,
\be
\sP^\mu={1\over i}\sum_{n=1}^N {\partial\over \partial \sx^\mu_n},\; \qquad
\sJ^{\mu\nu}={1\over i}\sum_{n=1}^N\left(\sx_n^\mu{\partial \over \partial \sx_n^\nu}
-\sx_n^\nu{\partial \over \partial
\sx_n^\mu}+\fourth[\gamma_{ne}^\mu,\gamma_{ne}^\nu]\right)\; .
\ee

In general Green operators are  realized  by $E(4)$ invariant  tempered distributions
$G_N(\sx_1,\dots \sx_N;\sy_N,\dots,\sy_1)$. The cluster  properties
can be conveniently realized using formal annihilation  operators 
$a(\sx),\; b(\sx)$ and creation operators  $a^\dagger(\sx),\; b^\dagger(\sx)$ which
satisfy the commutation
relations\cite{coester54}
\bea
\{a(\sx),a^\dagger(\sy)\} &=& \delta^{(4)}(\sx-\sy)\; ,\qquad 
\{b(\sx),b^\dagger(\sy)\} = \delta^{(4)}(\sx-\sy)\cr\cr
\{a(\sx),b(\sy)\} &=&\{a(\sx),b^\dagger(\sy)\} =0\; .
\eea 
The Green function of $N$ free particles is related to the single-particle
Green function $G(\sx-\sy)$ by the expression
\be
G_{0N}(\sx_1,\dots,\sx_n;\sy_n,\dots,\sy_1)=
\bra{0}a(\sx_1)\cdots a(\sx_N)b(\sy_N)\cdots b(\sy_1)e^{\bf S_1}\ket{0}\; .
\ee
where 
\be
{\bf S_1}:= \int d^4\sx \int d^4\sy a^\dagger(\sx)b^\dagger(\sy)G(\sx-\sy).
\ee
and $\ket{0}$ is the  cyclic vector that is annihilated by $a(\sx)$, 
$b(\sx)$.
With the definition 
\bea
 \psi(\sx)&:=&e^{-\bf S_1}a(\sx)e^{\bf S_1}= a(\sx) -\int d^{(4)}\sy
G(\sx-\sy)b^\dagger(\sy)\; , \cr\cr
\bar\psi(\sx)&:=& e^{-\bf S_1}b(\sx)e^{\bf S_1}= b(\sx) +\int d^{(4)}\sy
a^\dagger(y)G(\sy-\sx)
\eea
it follows that
\be
G_{0N}=\bra{0}\psi(\sx_1)\cdots \psi(\sx_N)\bar\psi(\sy_N)\cdots
\bar\psi(\sy_1)\ket{0}\; .
\ee
The cluster structure of the general $N$-particle Green functions is realized 
by the expression\cite{coester58}
\be
G_N(\sx_1,\,\dots\,,\sx_N;\sy_N,\,\dots\,
,\sy_1)=\<0|a(\sx_1)\cdots a(\sx_N)b(\sy_N)\cdots b(\sy_1)\exp \left(\sum_n {\bf
S_n}\right)|0\>\; ,
\ee
if the functions $S_n(\sx_1,\dots,\sx_n;\sy_n,\dots,\sy_1)$ in 
\bea
{\bf S_n}&:=&{1\over {n!}^2}\int d^4\sx_1\dots \int d^4\sy_1\dots \cr\cr &\times&
a^\dagger(\sx_n)\cdots a^\dagger(\sx_1)b^\dagger(\sy_1)\cdots b^\dagger(\sy_n)
S_n(\sx_1,\,\dots\,,\sx_n;\sy_n,\,\dots\,,\sy_1)
\eea
vanish for separation of the points into widely separated clusters.
The cluster structure of the the Green operator is realized imposing
this structure on the inverse Green operator,
\be
{\bf G}^{-1} = \exp\left[\int d^4 \sx \int d^4 \sy a^\dagger(\sx)b^\dagger(\sy)
S^{-1}(\sx-\sy) +\sum_{n\geq 2} {\bf U}_n\right]\; .
\ee
The interaction operators ${\bf U}_n$ for different $n$ are independent. In
particular we may assume  ${\bf U}_n\equiv 0$ for $n>2$.
\be
{\bf U}_2=\fourth \int d^4\sx_1 \int d^4\sx_2\int d^4\sy_1\int d^4 \sy_2
a^\dagger(\sx_1)a^\dagger(\sx_2)b^\dagger(\sy_2)b^\dagger(\sy_1)
U(\sx_1,\sx_2;\sy_2,\sy_1)\; ,
\ee
where $U(\sx_1,\sx_2;\sy_2,\sy_1)$  may  for instance be given by eq. (\ref{BSU}).

The full Green function satisfying cluster separability is then given  for any
$N$ by
\be
G(\sx_1,\dots,\sx_N;\sy_N,\dots,\sy_1)= 
 \bra{0}\psi(\sx_1)\cdots \psi(\sx_n)\bar\psi(\sy_n)\cdots 
\bar \psi(\sy_1)e^{{\bf A}}\ket{0}\; ,
\ee
with
\be
{\bf A}:= \fourth \int d^4\sx_1 \int d^4\sx_2\int d^4\sy_1\int d^4 \sy_2
\psi(\sx_1)\psi(\sx_2)\bar \psi^\dagger(\sy_2)\bar\psi^\dagger(\sy_1)
U(\sx_1,\sx_2;\sy_2,\sy_1)\; .
\ee
\section{Conclusions}

Kinematic Poincar\'e covariance of state vectors of many-body systems requires
a dynamically determined semi-definite inner product measure. An effective
realization is based on an auxiliary Hilbert space endowed with a unitary
representation of the four-dimensional Euclidean group. A self-adjoint unitary
involution operator provides a Poincar\'e invariant indefinite inner product.
The Euclidean invariant Green operator specifies the Poincar\'e invariant
semi-definite inner product of the subspace of physical states. In this framework
two-body Green operators are sufficient to determine many-body Green functions
satisfying cluster separability.

\section*{Acknowledgments}
This work  was supported in part by the Department of Energy, Nuclear
Physics Division, under contracts W-31-109-ENG-38 and DE-FG02-86ER40286.

\end{document}